\documentclass[preprint,12pt]{article}
\usepackage[margin=2.5cm]{geometry}
\usepackage[round]{natbib}
\usepackage{bm}
\usepackage{float}
\usepackage{graphicx}
\usepackage{import}
\graphicspath{ {./Images/} }
\usepackage{xcolor}
\usepackage{layouts}
\usepackage{amsmath} 
\usepackage{algorithm}
\usepackage{algpseudocode}
\usepackage{siunitx} 
\usepackage{caption, subcaption}
\usepackage{pgfplots}
\usepackage{newfloat}
\usepackage{tabularx}
\usepackage{amssymb}
\usepackage{afterpage}

\newcommand{\cs}{c_\text{s}}

\title{Parametric 3D Convolutional Autoencoder for the Prediction of Flow Fields in a Bed Configuration of Hot Particles}
\author{
    Ali Mjalled\textsuperscript{1},
    Reza Namdar\textsuperscript{2},
    Lucas Reineking\textsuperscript{1},
    \\[0.5ex]
    Mohammad Norouzi\textsuperscript{2},
    Fathollah Varnik\textsuperscript{2},
    Martin M{\"o}nnigmann\textsuperscript{1}
}
\date{}

\begin{document}

\maketitle

\begin{center}
\small
\textsuperscript{1} Automatic Control and Systems Theory, Ruhr-Universit{\"a}t Bochum, Universit{\"a}tstraße 150, Bochum, 44801, Germany\\
\textsuperscript{2} Interdisciplinary Centre for Advanced Materials Simulation, Ruhr-Universit{\"a}t Bochum, Universit{\"a}tstraße 150, Bochum, 44801, Germany 
\end{center}

\begin{abstract}
The use of deep learning methods for modeling fluid flow has drawn a lot of attention in the past few years.
Here we present a data-driven reduced model for predicting flow fields in a bed configuration of hot particles. 
The reduced model consists of a parametric 3D convolutional autoencoder.
The neural network architecture comprises two main components.
The first part resolves the spatial and temporal dependencies present in the input sequence, while the second part of the architecture is responsible for predicting the solution at the subsequent timestep based on the information gathered from the preceding part.
We also propose the utilization of a post-processing non-trainable output layer following the decoding path to incorporate the physical knowledge, e.g., no-slip condition, into the prediction.
The reduced model is evaluated by comparing the predicted solutions of the reduced model with the high-fidelity counterparts.
In addition, proper orthogonal decomposition is employed to systematically analyze and compare the dominant structures present in both sets of solutions.
The assessment of the reduced model for a bed configuration with variable particle temperature showed accurate results at a fraction of the computational cost required by traditional numerical simulation methods.
\end{abstract}

\textbf{Keywords:} Reduced model, Lattice Boltzmann, Neural networks, Autoencoder

\section{Introduction}
Over the past few decades, numerous experimental and numerical studies have been conducted to understand the flow behavior and heat transfer of fluids around solid bodies.
Within this broad spectrum of research, great emphasis has been given to flow fields in packed beds, where the transfer of heat, mass and momentum are increased by providing an extended interfacial surface between the fluid and solid bodies.
Packed beds play a crucial role in a variety of areas, ranging from chemical processing such as separation \citep{Bader2015}, absorption \citep{Lin2003} and catalytic processes \citep{AMBROSETTI2020} to energy storage applications like sensible heat storage systems \citep{ELOUALI2019}, and advanced adiabatic compressed air energy storage \citep{GEISSBUHLER2018}.
This wide range of applications as well as the effect of transfer phenomena on process efficiency, product quality and environmental sustainability \citep{bird2002}, motivate investigating the key features of packed beds such as flow field, pressure drop, packing material, bed height, and packing density to ensure optimal performance.  

Numerical methods are one of the most effective tools for analyzing the behavior of packed beds and optimizing their design parameters.
The discrete element method (DEM), computational fluid dynamics (CFD) and the lattice Boltzmann method (LBM) are some of the common approaches that have been used to simulate packed bed configurations.
DEM simulations have been applied to examine both the geometrical features of the packed beds and the movement of particles within them.
For example, \cite{KOU2020} used DEM to examine the effect of shape, restitution coefficient, sliding coefficient, rolling coefficient, and diameter ratio of cubical particles on the behavior of the packed beds.
CFD simulations, on the other hand, can capture the detailed flow field within the bed and the interaction between the fluid and particles.
\cite{KUTSCHERAUER2022} investigated flow field and heat transfer for a slender packed bed under various conditions.
Moreover, a combination of different methods can be adopted. 
\cite{Bai2009} used a coupled DEM-CFD model to study flow field and pressure drop in fixed bed reactors.
LBM, which is a relatively a new numerical method, has been utilized to simulate the interactions between fluid and particles, as well as the chemical reactions occurring within packed beds.
\cite{QI2019}, for example, used LBM to investigate non-Newtonian fluid flow through a packed bed of uniform spheres. 

The aforementioned numerical methods offer a more efficient and cost-effective alternative to experimental analysis for modeling diverse, packed bed configurations with varying properties.
However, simulating large-scale packed beds using these methods requires substantial computational resources, making it a time-consuming and expensive process.
Even with highly parallelizable techniques such as LBM, the simulation of large and complex packed beds can be limited, particularly when it comes to optimization problems.

In this context, building reduced models for packed bed simulations is an essential step toward reducing the computational cost and allowing for sensitivity analysis.
The majority of model reduction approaches rely on mapping the high dimensional space into a smaller one on which the reduced model is defined.
One of the most commonly employed methods in this context is based on proper orthogonal decomposition (POD) followed by a Galerkin projection (see, e.g., \cite{Benner2015}, for a survey, and \cite{Reineking2023a} for an engineering application).
This method uses POD to extract a set of dominant modes that serve as a basis for the reduced space. Subsequently, Galerkin projection is employed to extract a small number (order of ten) of ordinary differential equations by projecting the governing equations into the reduced space.
However, the POD-Galerkin method requires access to the governing equations of the model, which may not be available in many cases.
Therefore, significant work has been dedicated to mitigate this dependency by developing complete data-driven reduced modeling frameworks using artificial intelligence (e.g., \cite{Pawar2019, Wang2019}).
Many of the methods developed rely on replacing the Galerkin projection step with a deep neural network that predicts the temporal evolution of the coefficients associated with the POD modes. 
In light of this, various architectures have been used, including deep feedforward (e.g., \cite{Mjalled2023a, Lui_2019}), convolution (e.g., \cite{Wu2020}) and recurrent (e.g., \cite{Akbari_2022}) neural networks, to name just a few.

Despite the advantage of POD as an effective tool to reduce the dimensionality of data, its linear basis approximation makes it incompatible with highly nonlinear systems where a large number of modes is required to capture the most dominant energy content of the data.
Large numbers of modes are required because nonlinear phenomena are approximated as a linear combination of the POD modes.
On the other hand, there are several available approaches to learning nonlinear manifolds, such as Kernel PCA \citep{KPCA} and Laplacian eigenmaps \citep{LapEigen}. 
These techniques, however, do not provide an analytical relationship to reconstruct the original data from the compressed representation.
An alternative approach consists of using autoencoders (AEs) as a tool for learning nonlinear manifolds (e.g., \cite{Fu2023}). AEs not only offer the ability to learn intricate relationships within the data but also provide a structured means for reconstructing the high-dimensional representation. 

AEs are a special type of unsupervised neural network that are trained to learn hidden features from an input. 
Their architecture consists of two parts, namely, the encoder and the decoder.
The former compresses the high-dimensional input into a low-dimensional representation, while the latter learns how to reconstruct the input.
AEs have been extensively used to build data-driven reduced models of flow fields around solid bodies. 
\cite{Fukami_2021} investigated the ability of AEs for dimensionality reduction of flow field data in the wake of a two-dimensional cylinder.
However, his work focused only on data compression and did not predict temporal evolution.
In contrast, \cite{Xu2020} developed a multi-level AE reduced model that addresses all aspects of the flow, including dimensionality reduction, temporal evolution, and parameter variation. 
\cite{Hou2022} developed a hybrid AE-LSTM reduced model for the flow field prediction around submarines, where the LSTM (long short-term memory) layer has been utilized to resolve the dynamics of the compressed space.
\cite{Hesgawa2020} proposed a method to construct AE-based reduced models for unsteady flows around bluff bodies of various shapes.
Apart from AEs, other neural network architectures have been used to approximate flow fields around different shapes, such as airfoils \citep{Sekar2019} and automobiles \citep{Guo2016}.

Different variations of AEs have been used for reduced flow field modeling.
In their pioneering work, \cite{Pant2021} developed a reduced model framework based on 3D convolutional AE for the temporal evolution of fluid simulation.
The model demonstrated accurate results across various simulations, including the flow field prediction around a 2D circular cylinder.
In this work, we build upon their findings and extend the framework by making it parametric. 
This modification empowers the model to not only reproduce known simulations but also to predict the flow field corresponding to a new simulation parameter.
We apply the extended framework to a bed configuration of hot particles, wherein we propose the usage of a post-processing non-trainable output layer to encompass the physical knowledge of the system into the prediction, e.g., no-slip condition at the fluid-solid interface.

The remainder of the paper is organized as follows. Sec.~\ref{sec:num_model} presents the numerical model for generating the simulation database. A comprehensive overview of AEs is given in Sec.~\ref{sec:autoencoders}. Sec.~\ref{sec:RM} explains the framework used in this work to build a parametric data-driven reduced model based on 3D convolutional AE. The results are presented and discussed in Sec.~\ref{sec:results}. Finally, conclusions are provided in Sec.~\ref{sec:conclusion}.

\section{Numerical model}
\label{sec:num_model}
We present in this section the governing equations together with a brief explanation of the numerical model used in this study.
The model is validated by comparing the simulation results with benchmark data.
The results of the numerical model will serve as training data for the reduced model.
\subsection{Governing equations and lattice Boltzmann-finite difference model}
A variation of the Navier-Stokes equations is employed that accounts for the compressibility arising from the variation of temperature.
It should be noted that these equations hold true only in low Mach number flows.
The governing equations that ensure the conservation of mass, momentum and energy for this model are:
\begin{align}
\partial_t \rho + \bm{\nabla}\cdot (\rho \bm{u}) &= 0  \label{eq:mass-cons}\\
\partial_t (\rho \bm{u}) + \bm{\nabla}\cdot (\rho \bm{u} \otimes\bm{u}) &= - \bm{\nabla}P_h+ \bm{\nabla}\cdot \bm{\tau} + \bm{F} \label{eq:momentum-cons}\\
\partial_t  \rho h + \bm{\nabla}\cdot(\rho \bm{u} h) &= -\bm{\nabla} \cdot \bm{q}  + \partial_t P_t \label{eq:energy-cons}\\
\rho &= \frac{M_\text{w}\, P_t}{R\,T} \label{eq:rho-in-terms-of-P0-and-T},
\end{align} 
where $\rho$, $\bm{u}$, $\bm{\tau}$ and $\bm{F}$ represent the density, velocity, viscous stress tensor and body force per unit volume of the gas, respectively.
Moreover, $h$, $\bm{q}$, $R$, $M_\text{w}$ and $T$ stand for enthalpy, heat flux, universal gas constant (as the gas is assumed ideal in our simulation), molecular weight and temperature of the gas, respectively.
An important point to remark in this model is that pressure is divided into two parts, namely,  hydrodynamic pressure ($P_h$) and thermodynamic pressure ($P_t$).
$P_h$ considers the spatial and temporal variation of the pressure within the flow.
On the other side, $P_t$ is uniform spatially and depends only on time. 

To solve these equations, a combined lattice Boltzmann-finite difference (LB-FD) method was utilized.
In particular, the lattice Boltzmann method was employed to model the flow field, which takes into consideration heat expansion, while the energy equation was solved using the finite difference method.
The kinetic equations described below represent a modified version of the conventional lattice Boltzmann method that has the capability of recovering the thermal compressible form of Navier-Stokes equations beyond the Boussinesq approximation  (see, e.g., \cite{hosseini2019}, for more details):
\begin{align}
    \frac{\partial g_{\alpha}}{\partial t} &+ \bm{c}_\alpha \cdot \bm{\nabla}g_\alpha = -\frac{1}{\tau_r}(g_\alpha-g_\alpha^{\rm eq}) + \Xi_\alpha, \label{eq:lb-kin-eq}\\
    \Xi_\alpha &= (\bm{c}_\alpha - \bm{u}) \cdot [\bm{\nabla} {\rho \cs^2} (f^{\rm eq}/\rho - w_\alpha) + f^{\rm eq}/\rho] + {\rho \cs^2}w_\alpha \bm{\nabla} \cdot \bm{u}  \label{eq:lb-source-term},\\
    f_\alpha^{\rm eq} &= \rho w_\alpha\left(1+\frac{1}{\cs^2}\bm{c}_\alpha\cdot\bm{u}+\frac{1}{2\cs^4}{(\bm{c}_\alpha\cdot\bm{u})}^2 - \frac{1}{2\cs^2}\bm{u}^2\right) \label{eq:lb-feq},\\
  g_\alpha^{\rm eq} &= \cs^2 f_\alpha^{\rm eq} + w_\alpha (P_h-\rho \cs^2) \label{eq:lb-geq},
\end{align}
where $g_\alpha$, $\bm{c}_\alpha$, $\tau_r$, $w_\alpha$ and $f_\alpha$ are the modified distribution function, discrete velocities, relaxation coefficient, weights, and the standard distribution function in the lattice Boltzmann model, respectively.
Sutherland's law is used to determine the dynamic viscosity $\mu$, which is then used to compute the relaxation coefficients $\tau_r$ according to
\begin{equation}
    \label{eq:relaxation coefficients}
    \mu=\rho \cs^2(\tau_r-0.5).
\end{equation}
The reader is referred to \cite{namdar2023} for more details about the numerical model. 

\subsection{Validation of the numerical model}
To ensure the validity of the numerical model, two tests were conducted and their corresponding results are compared against literature data.
These tests involved simulating gas flow around a single 2D square or circular body at a Reynolds number of 100.
Known velocity and pressure conditions were applied at inlet and outlet boundaries, respectively. The no-slip boundary condition was also applied on the fluid-solid contact while upper and lower boundaries were subjected to periodic boundary conditions.
To evaluate the momentum exchange, the drag coefficient $C_D$ was calculated for the isothermal case employing different grid resolutions.
We present the obtained results in Tab.~\ref{tab:drag-coeff} and compare them to benchmark data. The results demonstrate the convergence of the mesh starting from $\Delta x = \SI{3e-04}{}\mathrm{m}$, and an agreement with the benchmark data is obtained within a tolerable deviation of $9\%$.  
\begin{table}[]
\centering
\caption{Drag coefficient of square/circular body placed in a cross-flow at Reynolds number 100 under isothermal conditions. $\Delta x$ is the distance between two adjacent grid points in both spatial directions.}
\label{tab:drag-coeff}
\begin{center}
\begin{tabularx}{\textwidth}{|>{\centering\arraybackslash}X|>{\centering\arraybackslash}X|>{\centering\arraybackslash}X|>{\centering\arraybackslash}X|>{\centering\arraybackslash}X|}
\hline
 & $\Delta x=\SI{4e-04}{} \mathrm{m}$ & $\Delta x= \SI{3e-04}{} \mathrm{m}$ & $\Delta x= \SI{2e-04}{} \mathrm{m}$  & Benchmark data \\
\hline
square & $1.556$ & $1.549$ & $1.548$ & $1.43$ \citep{sohankar1997}\\
\hline
circular & $1.520$ & $1.517$ & $1.516$ & $1.39$ \citep{braza}\\
\hline
\end{tabularx}
\end{center}
\end{table}

\begin{table}[]
\centering
\caption{Average Nusselt numbers of square/circular body placed in a cross-flow at Reynolds number 100. The temperatures of gas and solid bodies are 300K and 400K, respectively.}
\label{tab:nusselt}
\begin{tabularx}{\textwidth}{|X|X|X|X|X|}
\hline
 & $\Delta x=\SI{4e-04}{} \mathrm{m}$ & $\Delta x= \SI{3e-04}{} \mathrm{m}$ & $\Delta x= \SI{2e-04}{} \mathrm{m}$  & Benchmark data \\
\hline
square & $4.217$ & $4.136$ & $4.135$ & $4.03$ \citep{sharma2004}\\
\hline
circular & $5.365$ & $5.314$ & $5.311$ & $5.23$ \citep{sharma2004}\\
\hline
\end{tabularx}
\end{table}

To validate the heat transfer rate, a similar analysis is performed to compute the Nusselt number \textit{Nu} for a flow passing over square/circular body with 100K temperature difference.
The calculation of the local \textit{Nu} and average $\overline{\textit{Nu}}$ is according to the definitions given by \cite{vierendeels2003}: 
\begin{align}
    \textit{Nu} &=\frac{D}{\lambda_0\, (T_\text{particle}-T_\text{inlet})} (\lambda \frac{\partial T}{\partial n})_\text{surf.}, \label{eq:loc-nu}\\
    \overline {\textit{Nu}} &= \frac{1}{A}\int_{A} \textit{Nu} \,\textit{dA} \label{eq:avg-nu},
\end{align}
where $\lambda$ and $\lambda_0$ are the heat conductivity of gas evaluated at temperature of the particle surface and mean temperature $T_0=(T_\text{particle}+T_\text{inlet})/2$, respectively.
As the Prandtl number \textit{Pr} is assumed to be constant, the heat conductivity is computed as $\lambda=(\mu c_p)/\textit{Pr}$.
The results depicted in Tab.~\ref{tab:nusselt} show an agreement between the obtained $\overline {\textit{Nu}}$ and the benchmark results within an accuracy of $2\%$.

Based on the results reported in Tabs.~\ref{tab:drag-coeff} and \ref{tab:nusselt} for the flow around single particle, the validity of the solver is approved.

\subsection{Simulation setup}
\begin{figure}[]
    \centering
    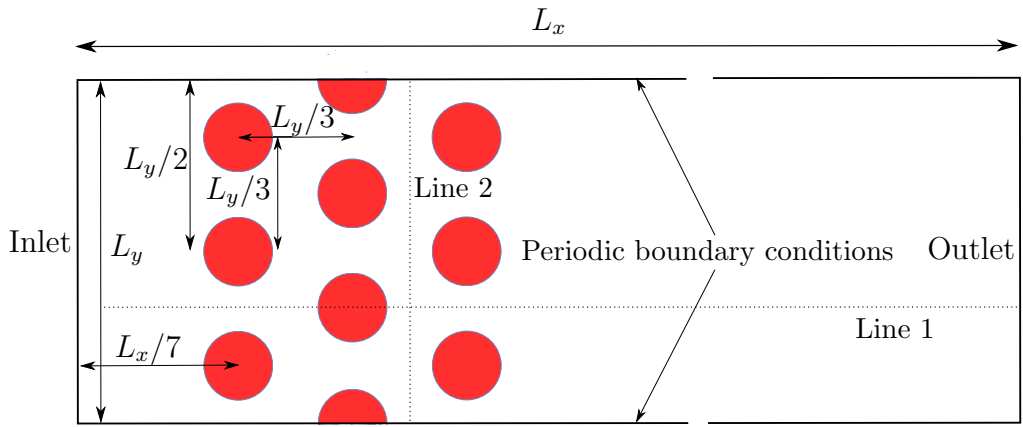
    \caption{Schematic representation of the 2D computational domain used to model the flow field and heat exchange in a bed configuration of hot particles. The dotted lines, denoted as Line 1 and Line 2, serve as demonstration lines for evaluating our model in Sec.~\ref{sec:results}.}
    \label{fig:main-config}
\end{figure}
The setup of the simulation is shown in Fig.~\ref{fig:main-config}.
The spatial dimension is 2D, wherein three rows of circular particles are arranged at a specific distance from each other.
The numerical domain is discretized using a cartesian grid of size $1800 \times 350$.
The corresponding simulation parameters are depicted in Tab.~\ref{tab:prop}.
This simple configuration is chosen in a first step to investigate the ability of deep learning to build reduced models for flow fields in packed bed configurations.
Despite the relatively sparse arrangement of particles, the implementation of periodic boundary conditions at the side boundaries of the bed helps to minimize surface effects and thus mimic bulk behavior to a certain degree.


\begin{table}
\begin{center}
    \caption{Parameters used to perform the simulations shown in Fig.~\ref{fig:main-config}.}
    \label{tab:prop}
    \begin{tabular}{ |p{3.2cm}|p{2.5cm}||p{3.2cm}|p{2.5cm}|  }
        \hline
        Parameter& Value &Parameter& Value\\
        \hline
        Fluid phase     & Air     &Solid phase&   Aluminium \\
        Time step (s)     & \SI{6.2e-07}{}     & Diameter (m)& \SI{3.0e-03}{}\\
        Grid size (m)    & \SI{4.3e-05}{}      & $T_\text{particle}$ (K)   &\SI{300}{} - \SI{1200}{}\\
        $Lx$ (m) & \SI{7.74e-02}{}  & $T_\text{inlet}$ (K)&  \SI{300}{}\\
        $Ly$ (m) & \SI{1.5e-02}{} & $P_\text{t,inlet}$ (Pa)&  \SI{1.0e+05}{}\\
        $c_{p, \text{air}}$ (\SI{}{\joule\per\kilogram \per\kelvin}) & \SI{1.0e+3}{}  & $M_{w, \text{air}}$ (\SI{}{\kilogram\per \mole}) & \SI{2.9e-02}{}\\
        Re &   200  & Pr \ & 0.71\\
        \hline
    \end{tabular}
\end{center}
\end{table}

\section{A brief introduction to autoencoders}
\label{sec:autoencoders}
\begin{figure}[H]
    \centering
    \includegraphics{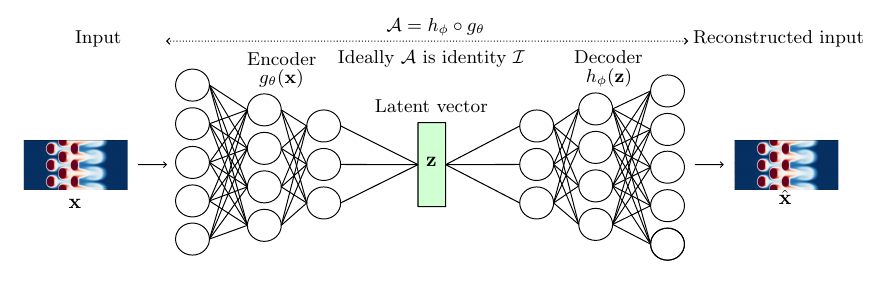}
    \caption{Example representation for a fully-connected autoencoder.}
    \label{fig:FCAE}
\end{figure}

AEs were first introduced by \cite{Rumelhart1986} as neural networks trained to reproduce their inputs. 
Unlike supervised learning, which relies on input-output pairs with explicit labels, training AEs is considered unsupervised learning because it does not require labeled data to be trained. Their objective is to learn an identity mapping $\mathcal{A}$.
In general, $\mathcal{A}$ can be identified using a neural network composed of two sub-networks, an encoder and a decoder.
The encoder function $g_{\theta}(\mathbf{x}):\mathbb{R}^N \rightarrow \mathbb{R}^p$ maps the input $\mathbf{x}$ into a lower dimensional representation $\mathbf{z}$, referred to as the latent vector, such that $p \ll N$.
The latent vector $\mathbf{z}$ is then passed to the decoder function $h_{\phi}(\mathbf{z}):\mathbb{R}^p \rightarrow \mathbb{R}^N$ which generates an approximated solution $\hat{\mathbf{x}}$ to the input.
The indices $\theta$ and $\phi$ refer to the training parameters of the encoder and decoder networks, respectively.
Training an AE consists of finding the optimal training parameters $\theta$ and $\phi$ that minimize a loss function $\mathcal{L}(\mathbf{x},\hat{\mathbf{x}})$ between the original and reconstructed inputs,i.e.,
\begin{equation}
    \label{eq:optim_AE}
     \theta_{\mathrm{opt}}, \phi_{\mathrm{opt}} = \mathrm{argmin}_{\theta, \phi}\mathcal{L}(\mathbf{x},h_\phi(g_\theta(\mathbf{x}))).
\end{equation}
By capturing the most important features and patterns of the input, the AE compresses the input into a condensed representation through its encoder function. This compression technique involves reducing the dimensionality of the input while maintaining critical information.
In this sense, AEs can be considered as a nonlinear generalization of POD.

The AE architecture can be varied depending on the data type and its input format.
Several types of AEs have been developed \citep{Zhai2018}, each tailored to address specific problems or exploit particular data features. 
In this paper, we will provide an overview of those relevant to our work.
\subsection{Fully-connected autoencoder}
This variant of AE uses fully-connected layers to find a compressed representation $\mathbf{z}$ for the input $\mathbf{x}$ (see Fig.~\ref{fig:FCAE}).
Every fully connected layer performs the following operation:
\begin{equation}
    \label{eq:FCL}
    \mathbf{y} = \mathbf{f}(\mathbf{W}\mathbf{x} + \mathbf{b}),
\end{equation}
where $\mathbf{y}\in \mathbb{R}^l$ is the output of the layer (input of the following layer in a deep network), $\mathbf{W} \in \mathbb{R}^{l \times N}$ is a weight matrix and $\mathbf{b} \in \mathbb{R}^l$ is the bias vector. 
The nonlinear activation function $\mathbf{f}:\mathbb{R}^l \rightarrow \mathbb{R}^l$ is applied element-wise to the output of the affine transformation.
This variation of AE is not suitable for handling high-resolution 2D input images due to its requirement of the flattening the input into a 1D vector.
In addition, flattening the image results in disregarding the spatial structure, i.e., the spatial connections between neighboring pixels are not explicitly maintained.
An alternative would be to use convolutional layers to preserve the spatial information.  

\subsection{Convolutional autoencoder}
\label{sec:conv}
\begin{figure}[h]
    \centering
    \includegraphics{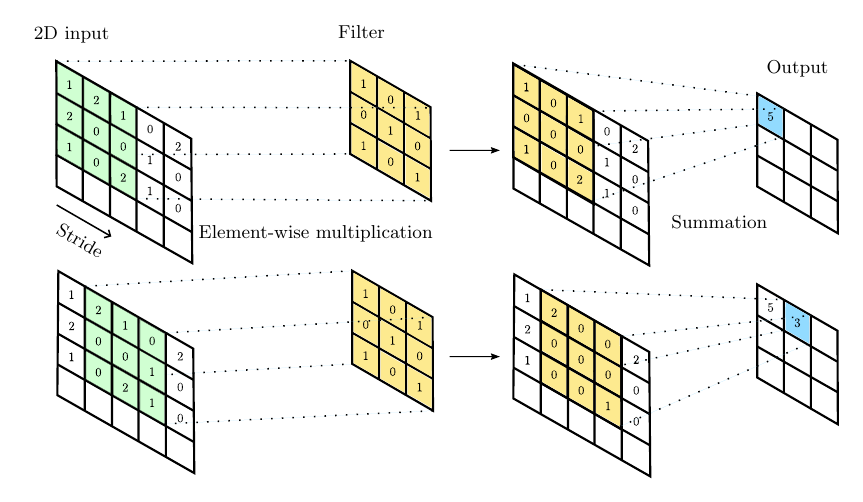}
    \caption{2D convolution operation with a $3\times 3$ filter and stride 1. Element-wise multiplication takes place between the filter and the corresponding entries of the input. The results are summed to calculate the output, followed by applying a nonlinear activation function in the way described in \eqref{eq:FCL}.}
    \label{fig:Conv}
\end{figure}
In this variation of AE, the fully connected layers are replaced by convolutional layers.
They are frequently used in computer vision tasks because they capture spatial dependencies effectively (see, e.g., \cite{Geng2016}).
As the name suggests, a convolutional layer applies a convolution operation to its input.
Convolution is the process of applying a kernel or filter to the input data and computing the element-wise multiplications between the filter entries and the corresponding input values.
The results are then summed.
Mathematically, the convolutional operation is expressed in the same way presented in \eqref{eq:FCL} with $\mathbf{W}$ being a circular matrix.
An illustration of a 2D convolutional operation is shown in Fig.~\ref{fig:Conv}. 
It should be noted that the underlying convolutional principle is the same for 1D and 3D cases.

\subsection{Temporal autoencoder}
Temporal AEs are used in many applications, e.g., video prediction and motion forecasting, where temporal coherence needs to be learned from sequential data (see, e.g., \cite{Zhao2017}).
The working principle of a temporal AE is slightly different from that of the conventional one described above. In a temporal AE, the focus is on the prediction of the temporal evolution of the input data,i.e.,
\begin{equation}
    \label{eq:TAE}
    \mathbf{x}(t+1) = \mathcal{T}(\mathbf{x}(t)),
\end{equation}
where $\mathbf{x}(t)$ and $\mathbf{x}(t+1)$ denote the input variable at the current and next timesteps, respectively, and $\mathcal{T}$ is the neural network that encloses the encoder and decoder functions.
Temporal AEs require supervised learning because the temporal dependency is learned by providing the ground truth sequence during training.
Both fully-connected and convolutional layers can be used to build the architecture of the temporal AE. 
The terminology autoencoder is used here to describe the underlying structure of an encoder-decoder model. 
Its use, however, departs from the conventional approach defined earlier, where the encoder compresses the input data, and the decoder reconstructs it at the output layer.
Instead, in a temporal AE, the encoder learns temporal relationships from a sequence, and the decoder reconstructs the data which serves as input at the following timestep (see Fig.~\ref{fig:framework}).

\section{Data-driven reduced model}
\label{sec:RM}
We extend the framework introduced by \cite{Pant2021} for constructing a neural network-based reduced model for fluid simulations by making it parametric, i.e., the dynamics are predicted for new values of the parameters that were not used during training.
The reduced model is represented as a 3D temporal convolutional AE.

\subsection{Model architecture}
We define the snapshot $\mathbf{s}(t_m) \in \mathbb{R}^{N_x \times N_y \times N_v}$ as a multichannel image of the flow field solution at time step $t_m$.
Each of the $N_v$ channels represents the solution of a flow field variable (velocity magnitude or temperature in our case) defined on a uniform 2D Cartesian grid of dimension $N_x \times N_y$.

\begin{figure}[]
    \centering
    \includegraphics{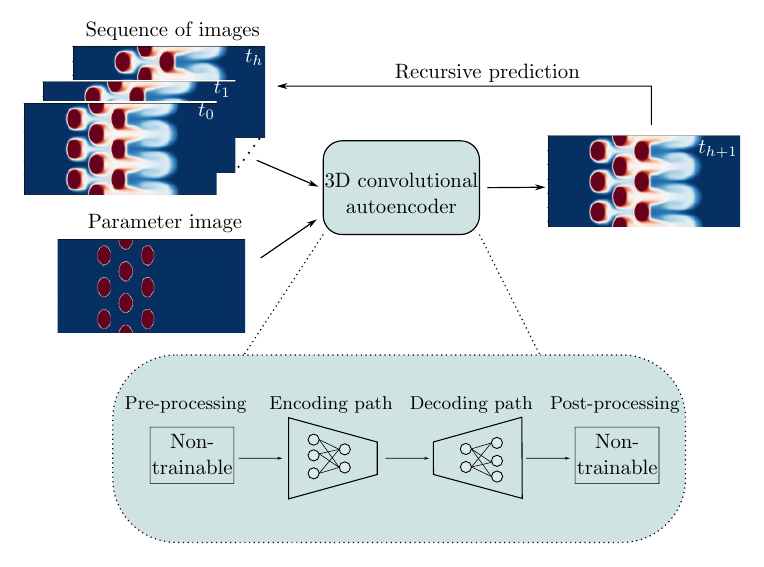}
    \caption{Data-driven framework for the parametric prediction of flow field simulations in packed bed configuration using 3D convolutional autoencoder.}
    \label{fig:framework}
\end{figure}

Dynamic flow field simulations can be analyzed by temporal sequence of solution snapshots. Given the previous $h$ snapshots and the corresponding simulation parameter value $\mu$, the reduced model predicts the flow field solution at the next time step, i.e., 
\begin{equation}
    \label{eq:data-driven rm}
    \mathbf{s}(t_{h+1}) \approx \mathcal{N}(\mu,\mathbf{s}(t_0),\cdots,\mathbf{s}(t_h)),
\end{equation}
where $\mathcal{N}$ is a nonlinear function that results from training a 3D temporal convolutional AE. 
As an extension of the 2D convolutional layers illustrated in Sec.~\ref{sec:conv}, the 3D convolutional layer used here not only operate on the spatial but also on the temporal dimension. 
Therefore, they are well suited for applications where both the spatial and temporal dimensions of the data are significant, such as video classification, action recognition, and spatio-temporal segmentation.
The parametric data-driven reduced model framework and the model architecture are presented in Fig.~\ref{fig:framework}. 
The 3D temporal convolutional AE used in this work employs an encoding-decoding path to learn the underlying dynamics of the spatio-temporal data.
The encoding path consists of four 3D convolutional layers with a tanh activation function.
A connection of non-trainable layers precedes the encoding path of our model to combine the different inputs to the model, i.e., the input sequence and the corresponding simulation parameter $\mu$, which is the temperature of the solid particles in our simulation.
It should be noted that $\mu$ is given to the network as an image of the initial condition of the flow.
The image format of the parameter value facilitates the combination of $\mu$ with the other inputs, as they are also presented in image form.
Once the temporal sequence has been analyzed in the encoding path, the neural network predicts the solution at the next time step by reconstructing the flow field image with the original size in the decoding path. 
For this purpose, four 2D convolutional transpose layers with tanh activation are used. 
The decoding path is followed by a non-trainable post-processing layer that effectively utilizes the physical information embedded within the parameter image. 
Specifically, we use the location of particles to enforce a no-slip condition to guide the prediction. 
Incorporating these constraints will improve the prediction accuracy of our framework. 
In order to improve the efficiency of the training, we use a batch normalization layer after every intermediate convolutional operation. This technique helps to speed up convergence and reduce overfitting \citep{batchnorm}. 
Tab.~\ref{tab:arch} shows the detailed architecture of the model. In this architecture, all the convolutional layers operate with a stride of 2 in all image dimensions to systematically decrease/increase the output size. The filter size used for the 3D convolutional layers is $2\times 4 \times 4$, while the 2D convolutional transpose layers use a filter of size $4 \times 4$.

\begin{table}
\caption{Detailed architecture of the 3D convolutional autoencoder. All convolutional and transpose convolutional layers use a stride of 2 in all image dimensions. The dimensions shown in this table correspond to $h=10$ and $N_x \times N_y = 256 \times 128$.}
\label{tab:arch}
\centering
\begin{tabularx}{\textwidth}{|X|X|X|}
            \hline
            Layer & Output Shape & Corresponding input(s) \\
            \hline
            Pre-processing & & \\
            \hline
            ParameterImage (Input) & (None, 256, 128) & [] \\
            ExpandLayer1 & (None, 256, 128, 1) & [ParameterImage] \\
            ExpandLayer2 & (None, 1, 256, 128, 1) & [ExpandLayer] \\
            SequenceImages (Input) & (None, 10, 256, 128, 2) & [] \\
            Tile & (None, 10, 256, 128, 1) & [ExpandLayer2] \\
            Concatenate & (None, 10, 256, 128, 3) & [SequenceImages, Tile] \\
            \hline
            Encoding path & & \\
            \hline
            Conv3D-1 & (None, 5, 128, 64, 64) & [Concatenate] \\
            Conv3D-2 & (None, 3, 64, 32, 128) & [Conv3D-1] \\
            BatchNorm-1 & (None, 3, 64, 32, 128) & [Conv3D-2] \\
            Conv3D-3 & (None, 2, 32, 16, 256) & [BatchNorm-1] \\
            BatchNorm-2 & (None, 2, 32, 16, 256) & [Conv3D-3] \\
            Conv3D-4 & (None, 1, 16, 8, 512) & [BatchNorm-2] \\
            BatchNorm-3 & (None, 1, 16, 8, 512) & [Conv3D-4] \\
            Reshape & (None, 16, 8, 512) & [BatchNorm-3] \\
            \hline
            Decoding path & & \\
            \hline
            Conv2D-transpose-1 & (None, 32, 16, 256) & [Reshape] \\
            BatchNorm-4 & (None, 32, 16, 256) & [Conv2D-transpose-1] \\
            Conv2D-transpose-2 & (None, 64, 32, 128) & [BatchNorm-4] \\
            BatchNorm-5 & (None, 64, 32, 128) & [Conv2D-transpose-2] \\
            Conv2D-transpose-3 & (None, 128, 64, 64) & [BatchNorm-5] \\
            BatchNorm-6 & (None, 128, 64, 64) & [Conv2D-transpose-3] \\
            Conv2D-transpose-4 & (None, 256, 128, 2) & [BatchNorm-6] \\
            \hline
            Post-processing & & \\
            \hline
            Output & (None, 256, 128, 2) & [ParameterImage, Conv2D-transpose-4] \\
            \hline
\end{tabularx}
\end{table}

\subsection{Input and output datasets}
The training data used in this study comprises high-dimensional simulation snapshots $\mathbf{s}(t_m)$ and their corresponding simulation parameter $\mu$. The high-dimensional snapshots are obtained from solving the numerical model described in Sec.~\ref{sec:num_model}, and to make them compatible with the input layer of the neural network, they are first interpolated on a uniform Cartesian grid of dimension $256 \times 128$.
The interpolated values are considered ground truth and will be used to train and evaluate our model.
Subsequently, the interpolated snapshots are normalized between -1 and 1 for numerical stability during the training process according to
\begin{equation}
    \label{eq:normalize}
    \chi_{\text{norm}} = 2\cdot\frac{\chi - \chi_{\text{min}}}{\chi_{\text{max}} -\chi_{\text{min}}} - 1, 
\end{equation}
where $\chi$ is the corresponding pixel value of the image and $\chi_{\text{min}}$ and $\chi_{\text{max}}$ are the minimum and maximum pixel values, respectively, identified for all snapshots for all training simulations parameterized with $\mu$.
It is important to note that each image channel, i.e., variable field, is scaled separately. The parameter images are also normalized as they are considered another input to the neural network. 

For the purpose of training, the snapshots are structured sequentially:
\begin{equation}
    X(\mu) = \begin{bmatrix}
            \mathbf{s}(t_0) & \mathbf{s}(t_1) & \cdots & \mathbf{s}(t_h) \\
            \mathbf{s}(t_1) & \mathbf{s}(t_2) & \cdots & \mathbf{s}(t_{h+1}) \\
            \vdots & \vdots & \cdots & \vdots \\
            \mathbf{s}(t_{N_t - h -1}) & \mathbf{s}(t_{N_t - h}) & \cdots & \mathbf{s}(t_{N_t - 1})
        \end{bmatrix}, \qquad
    Y(\mu) = \begin{bmatrix}
            \mathbf{s}(t_{h+1}) \\
            \mathbf{s}(t_{h+2}) \\
            \vdots \\
            \mathbf{s}(t_{N_t})
        \end{bmatrix},
\end{equation}
where $N_t$ is the total number of simulation snapshots, $X$ is the sequence of images given as input and $Y$ is the corresponding output for a single simulation.    

\subsection{Offline training and online prediction}
The supervised learning approach is used to approximate $\mathcal{N}$, which consists of finding the optimal training parameters $\theta$ that minimizes a loss function $\mathcal{L}(\mathbf{s}(t_{m}),\hat{\mathbf{s}}(t_{m}))$ between the ground truth $\mathbf{s}(t_{m})$ and the predicted solution $\hat{\mathbf{s}}(t_{m}))$
\begin{equation}
    \label{eq:optim NN}
    \theta_{\mathrm{opt}} = \mathrm{argmin}_{\theta}\mathcal{L}(\mathbf{s}(t_{m}),\hat{\mathbf{s}}(t_{m})).
\end{equation}
The optimization problem shown in \eqref{eq:optim NN} can be solved using the gradient descent algorithm or one of its variants, such as the Adam optimizer \citep{Kingma2014}. Backpropagation \citep{Rumelhart1986we} is used to compute the partial derivatives required to update the trainable parameters $\theta$. The optimization process continues until convergence or until a certain stopping criterion is met. The training procedure is summarized in Algorithm~\ref{alg:NN}.

\begin{algorithm}[t]
\caption{Offline training}\label{alg:NN}
\begin{algorithmic}[1]
\Require $\{ \mu^1, \cdots \mu^{N_s} \}, \mathrm{Epochs}, \mathrm{learning \ rate} = l_r$;
\For{$s=1,\cdots,N_s$}
    \State Build $X(\mu^s)$ and $Y(\mu^s)$;
    \State Add to the complete training inputs array $X \leftarrow X(\mu^s)$;
    \State Add to the complete training outputs array $Y \leftarrow Y(\mu^s)$;
    \State Add to the parameter array $\mu \leftarrow Y(\mu^s)$;
\EndFor
\State Initialize the trainable parameters $\theta$ randomly;
\For{$i=1,\cdots,\mathrm{Epochs}$}
\For{$j=1,\cdots,\mathrm{length}(X)$}
\State Feedforward in the network to make a prediction $\hat{\mathbf{s}} = \mathcal{N}(\mu[j], X[j])$;
\State Compute the loss $\mathcal{L}(Y[j],\hat{\mathbf{s}})$;
\State Compute the partial derivatives $\frac{\partial \mathcal{L}}{\partial \theta}$;
\State Update the trainable parameters using gradient descent algorithm $\theta \leftarrow \theta - l_r \frac{\partial \mathcal{L}}{\partial \theta}$;
\EndFor
\EndFor
\State \Return $\mathcal{N}$
\end{algorithmic}
\end{algorithm}
\begin{algorithm}[H]
\caption{Online prediction}\label{alg:prediction}
\begin{algorithmic}[1]
\Require $\mathcal{N}, \mu^{\text{val}}, h, \chi_{\text{min}}, \chi_{\text{max}}$;
\For{$m=0,\cdots,h$}
    \State Solve the numerical model for $\mathbf{s}(t_m)$;
    \State Scale the snapshot using \eqref{eq:normalize};
\EndFor
\For{$m=h,\cdots,N_t-1$}
\State make a prediction $\mathbf{s}(t_{m+1}) = \mathcal{N}(\mu^{\text{val}}, \mathbf{s}(t_m),\cdots,\mathbf{s}(t_{m-h}))$;
\State Unscale the prediction $\mathbf{s}^{\text{orig}}(t_{m+1}) = \frac{1}{2}(\mathbf{s}(t_{m+1})+1)(\chi_{\text{max}} - \chi_{\text{min}}) + \chi_{\text{min}}$;
\State Add to the prediction array $\mathbf{S}^{\text{orig}} \leftarrow \mathbf{s}^{\text{orig}}(t_{m+1})$;
\EndFor
\State \Return $\mathbf{S}^{\text{orig}}$
\end{algorithmic}
\end{algorithm}

Once the neural network has been trained, it can be used to predict the flow field solution for a new value of $\mu$ that was not used during training.
The prediction process starts by providing the first $h$ snapshots as an input to the model.
In this work, we use the ground truth snapshots to make our initial predictions. We emphasize that the prediction is recursive (see Fig.~\ref{fig:framework}). The online step is concluded with an unscaling process to revert the snapshots to their original scale. Algorithm~\ref{alg:prediction} summarizes the prediction of the flow field solution given a new value of the simulation parameter.

\section{Results}
\label{sec:results}
\begin{figure}[]
    \centering
    \includegraphics[width=\textwidth]{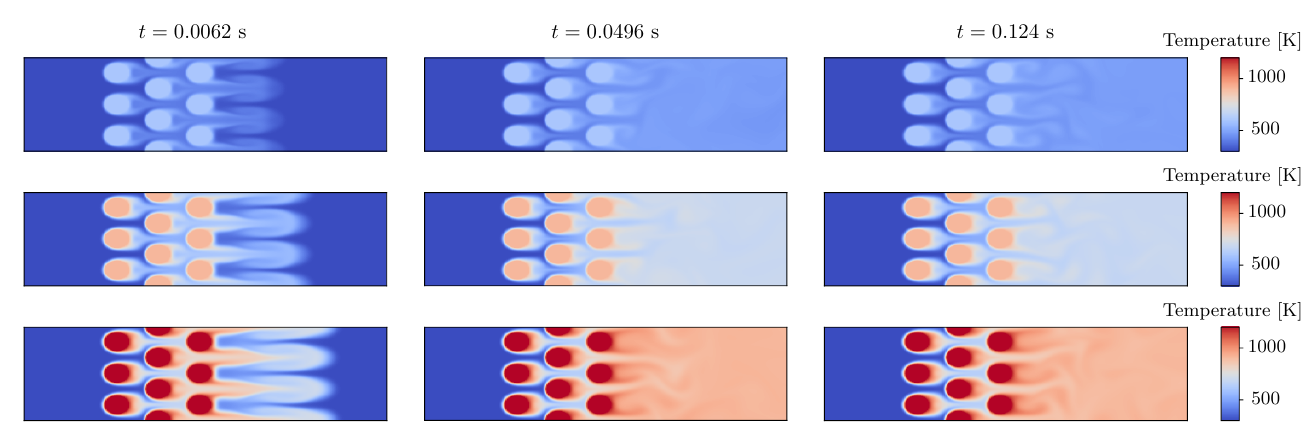}
    \caption{A comparison of temperature fields in snapshots obtained from simulations parametrized with (top) $T^{\text{par, tr}}=600$ K, (middle)  $T^{\text{par, tr}}=900$ K, and (bottom) $T^{\text{par, tr}}=1200$ K. The snapshots correspond to time steps $t=0.0062, \ 0.0496, \ 0.124$ s.}
    \label{fig:SnapshotsTemp}
\end{figure}
\begin{figure}[]
    \centering
    \includegraphics[width=\textwidth]{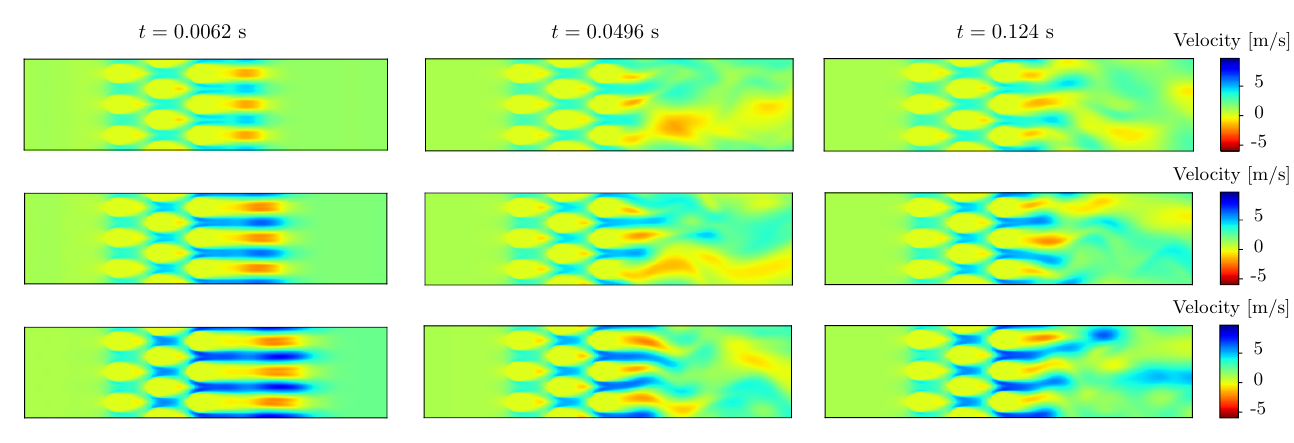}
    \caption{A comparison of velocity fields in snapshots obtained from simulations parametrized with (top) $T^{\text{par, tr}}=600$ K, (middle)  $T^{\text{par, tr}}=900$ K, and (bottom) $T^{\text{par, tr}}=1200$ K. The snapshots correspond to time steps $t=0.0062, \ 0.0496, \ 0.124$ s.}
    \label{fig:SnapshotVels}
\end{figure}
We present a comprehensive evaluation of the performance of the parametric 3D temporal convolutional AE in this section.
Our primary objective is to develop a data-driven framework for the high-dimensional dynamic system presented in Sec.~\ref{sec:num_model}.
The neural network consists of the 3D temporal convolutional AE explained in Sec.~\ref{sec:RM} that predicts the flow field solution at the next timestep given the previous $h=10$ snapshots and the corresponding value of $\mu$.
The length $h$ of the input sequence is a hyperparameter of the model and can be adjusted if needed.

\subsection{Comparative analysis between the high-fidelity and the reduced models}
The training and validation datasets were obtained by running the high-fidelity model presented in Sec.~\ref{sec:num_model}. Each simulation is parameterized by the temperature of the particles $T^{\mathrm{par}}$.
The neural network is trained using Algorithm~\ref{alg:NN} for the particle temperature parameters $ T^{\text{par, tr}} = 300,400,500,600,700,800,900,1000,1100,1200$ K.
A total number of 4000 snapshots (400 for every training simulation) were collected with a uniform timestep of $6.2\times 10^{-4}\text{ s}$.
Figures \ref{fig:SnapshotsTemp} and \ref{fig:SnapshotVels} provide a representative comparison of snapshots obtained from different training simulations, showcasing the temperature and velocity fields, respectively.
In the context of temperature fields in Fig.~\ref{fig:SnapshotsTemp}, the increase in particle temperatures is associated with more heat transfer between particles and the fluid medium.
Notably, consistent patterns in the transient solution can be observed across various simulations, albeit with varying amplitudes corresponding to particle temperatures.
This trend is also observed in the post-transient solution, with minimal variation among different simulations in the wake behind the cylinders of the last row.
On the other hand, the velocity field results presented in Fig.~\ref{fig:SnapshotVels} show that the solution structure is similar only during the transient phase, while the post-transient solution exhibits large structural variation between various simulations, particularly in the downstream region.
In addition, across all training simulations, it is observed that the velocity field does not converge to a limit cycle. 
\begin{figure}[t]
    \centering
    \includegraphics[width=\textwidth]{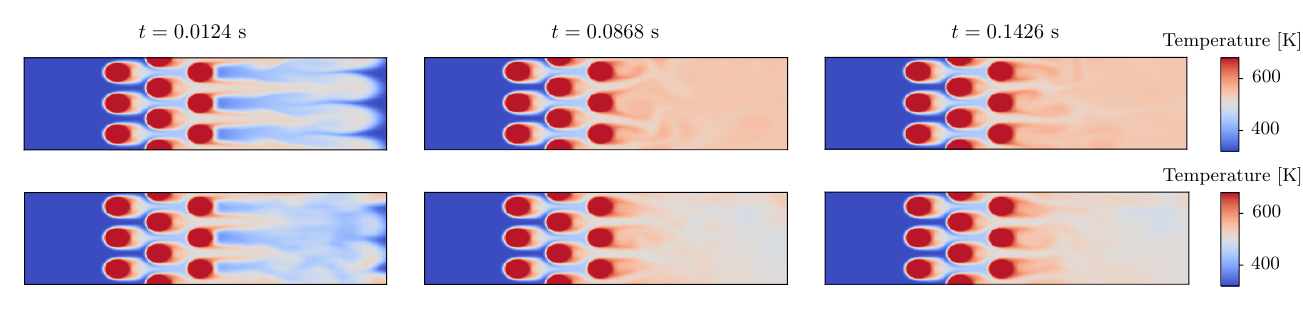}
    \caption{Comparison of the temperature fields obtained from (top) lattice Boltzmann method and (bottom) 3D convolutional autoencoder at time levels $t=0.0124, \ 0.0868, \ 0.1426 \text{ s}$. The results correspond to a validation simulation parameterized with $T^{\text{par, val}}=750$ K.}
    \label{fig:temp_plots}
\end{figure}
\begin{figure}[t]
    \centering
    \includegraphics[width=\textwidth]{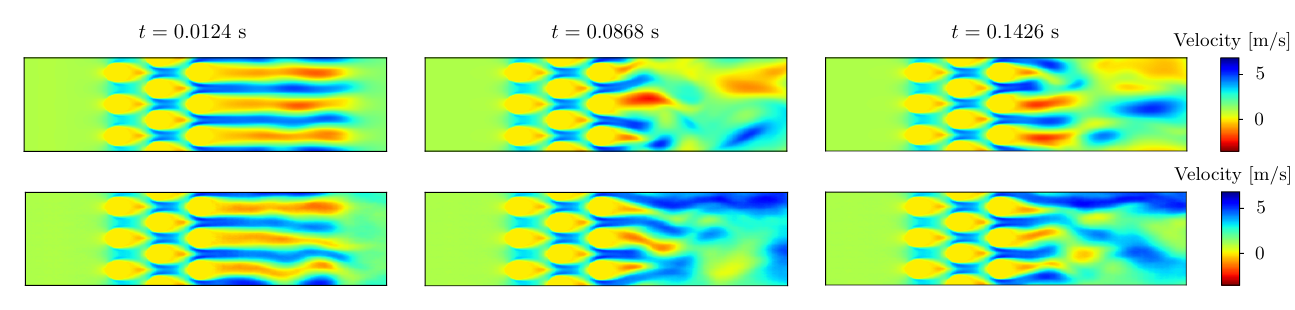}
    \caption{Comparison of the magnitude of the velocity fields obtained from (top) lattice Boltzmann method and (bottom) 3D convolutional autoencoder at time levels $t=0.0124, \ 0.0868, \ 0.1426 \text{ s}$. The results correspond to a validation simulation parameterized with $T^{\text{par, val}}=750$ K.}
    \label{fig:vel_plots}
\end{figure}

\begin{figure}[h]
    \centering
    \includegraphics[width=\textwidth]{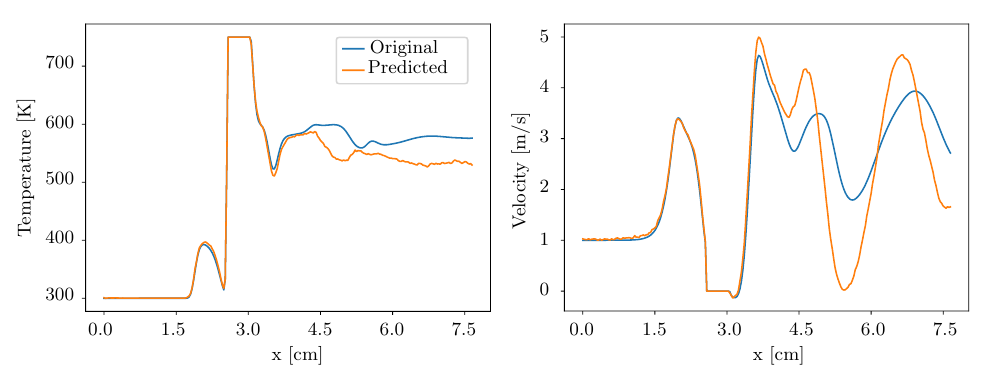}
    \caption{Comparison between (blue) original and (orange) predicted solutions for the temperature and velocity along Line 1, illustrated in Fig.~\ref{fig:main-config}, at time $t=0.1426\text{ s}$. The results correspond to a validation simulation parameterized with $T^{\text{par, val}}=750$ K.}
    \label{fig:line_eval}
\end{figure}
\begin{figure}[h]
    \centering
    \includegraphics[width=\textwidth]{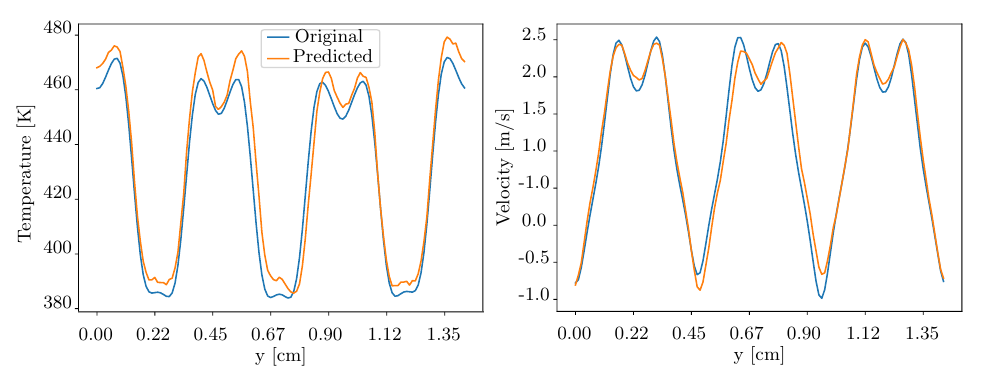}
    \caption{Comparison between (blue) original and (orange) predicted solutions for temperature and velocity along Line 2, illustrated in Fig.~\ref{fig:main-config}, at time $t=0.093\text{ s}$. The results correspond to a validation simulation parameterized with $T^{\text{par, val}}=550$ K.}
    \label{fig:line_eval_vert}
\end{figure}

In the training phase of the reduced model, we employ the Adam optimizer \citep{Kingma2014} with a learning rate of $\eta = 0.001$ and a batch size of $N_b = 10$ to find the optimal training parameters $\theta_{\mathrm{opt}}$.
The loss function used is the mean squared error, and the neural network is trained for $250$ epochs using the open-source deep learning library TensorFlow \citep{Abadi2016}.
Once the model is trained, we use Algorithm~\ref{alg:prediction} to predict the flow field solution for a new value of $\mu$.
Figures \ref{fig:temp_plots} and \ref{fig:vel_plots} evaluate the parametric 3D convolutional autoencoder model for the validation parameter $T^{\text{par, val}} = 750$ K.
The snapshots shown are chosen randomly and correspond to the solutions at $t=0.0124, \ 0.0868, \ 0.1426 \text{ s}$ for the temperature field (Fig.~\ref{fig:temp_plots}) and velocity field (Fig.~\ref{fig:vel_plots}).
It is evident that the first half of the domain, i.e., the interstitial flow between particles, was accurately predicted by our model with a very small visual difference in both temperature and velocity fields.
However, as we move downstream, a decline in the prediction accuracy becomes more visible in the temperature field, and to a greater extent in the velocity field.
Specifically, only the overall vortex shedding pattern is predicted in the downstream region without matching the spatial details.
These findings reflect the properties of the available training dataset and the differences highlighted in Figs.~\ref{fig:SnapshotsTemp} and \ref{fig:SnapshotVels}, where the effect of parameter variation on the temperature field is well represented by the training simulations, but less effectively conveyed in the velocity field within the downstream region.
\begin{figure}[]
    \centering
    \includegraphics[width=0.9\textwidth]{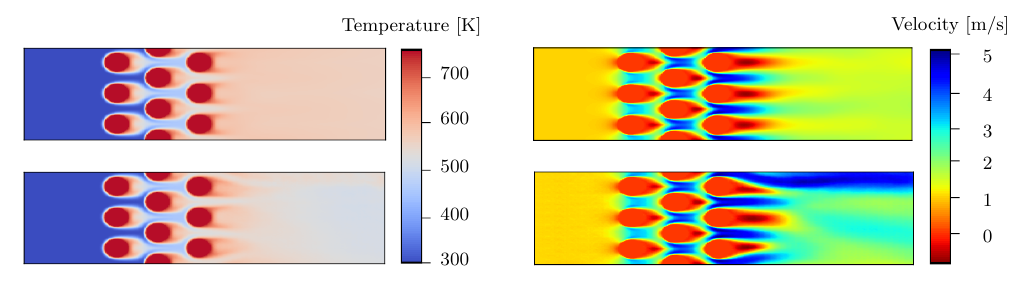}
    \caption{Temporal average fields obtained from (top) lattice Boltzmann method and (bottom) 3D convolutional autoencoder for the simulation parametrized with $T^{\text{par, val}} = 750$ K.}
    \label{fig:means}
\end{figure}
\begin{figure}[h]
    \centering
    \includegraphics[width=0.9\textwidth]{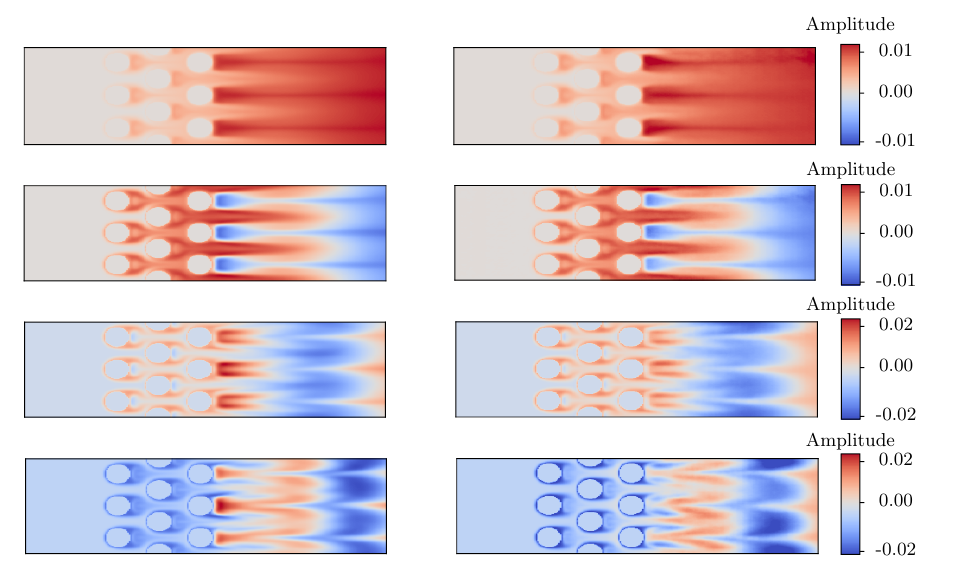}
    \caption{The four most dominant temperature modes, arranged from top to bottom. The left column corresponds to simulation results obtained from the lattice Boltzmann method, and the right column corresponds to results predicted using the 3D convolutional autoencoder for the simulation parametrized with $T^{\text{par, val}} = 750$ K.}
    \label{fig:modestemp}
\end{figure}
\begin{figure}[h]
    \centering
    \includegraphics[width=\textwidth]{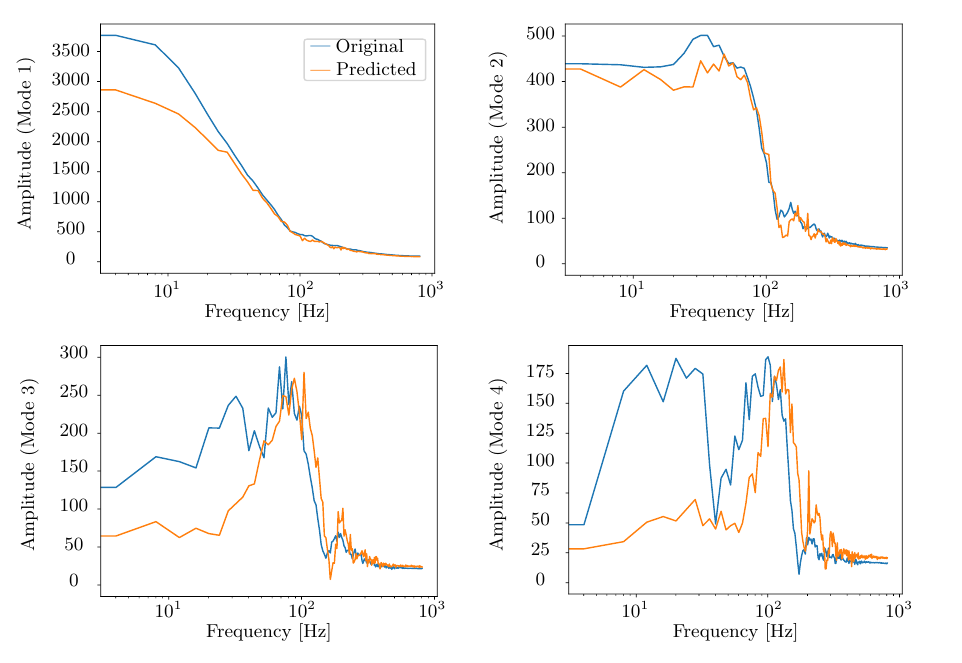}
    \caption{Power spectra for the temporal coefficients associated with the four most dominant POD temperature modes.}
    \label{fig:spectrumstemp}
\end{figure}

For the sake of a more quantitative comparison, the variations of the temperature and velocity along Line 1 and Line 2 from Fig.~\ref{fig:main-config} are plotted in Figs.~\ref{fig:line_eval} and \ref{fig:line_eval_vert}.
These plots, generated at randomly chosen time steps, correspond to two validation simulations parameterized with $T^{\text{par, val}} = 550$ K and $T^{\text{par, val}} = 750$ K.
Results demonstrate the capability of the parametric 3D convolutional AE to predict the flow field solution in a bed configuration of particles.
However, a noticeable deviation from the original solution is evident in the downstream domain where the flow becomes chaotic.
This deviation is attributed to the limited number of training simulations and the recursive prediction property of the model, which render accurate flow field prediction in chaotic regions a highly challenging task.
It is important to emphasize, however, that the model sufficiently captures the main flow pattern of vortex shedding, which is the dominant flow behavior in this region.

\subsection{Structural Comparison using proper orthogonal decomposition}
We pursue our evaluation of the reduced model by comparing the full original and predicted simulations for the validation parameter $T^{\text{par, val}} = 750$ K.
For this purpose, we employ POD (see Appendix.~A for a brief introduction to POD) as a means to extract and compare the dominant structures in the temperature and velocity fields of both results.
In the beginning, we decompose the quantities of interest into mean and fluctuation parts.
The corresponding temporal average fields are shown in Fig.~\ref{fig:means}.
A good agreement is observed between the averages obtained from the lattice Boltzmann method and the 3D convolutional autoencoder in the temperature field.
However, the predicted velocity in the downstream region is, on average, higher than the original one.
Remarkably, the greatest difference is located near to the upper boundary, where a high-velocity stream is predicted.
The four most dominant POD temperature modes of the predicted simulation are compared with the original ones in Fig.~\ref{fig:modestemp}.
The results demonstrate strong agreement between the two sets of modes, with a slight discrepancy appearing in higher modes.
Subsequently, we illustrate in Fig.~\ref{fig:spectrumstemp} the power spectra for the temporal coefficients associated with each of the POD temperature modes presented.
It should be noted that a one-to-one comparison is only meaningful in our case due to the highly similar structure of the POD temperature subspaces.
The comparison of spectra reveals a notable agreement in the high-frequency domain, suggesting consistent short-term dynamics across both the original and predicted simulations.
However, the congruent patterns in the low-frequency region point to slightly different long-term dynamics.
Additionally, it is observed that the two spectra deviate as we progress to higher modes.

\begin{figure}[h]
    \centering
    \includegraphics[width=0.9\textwidth]{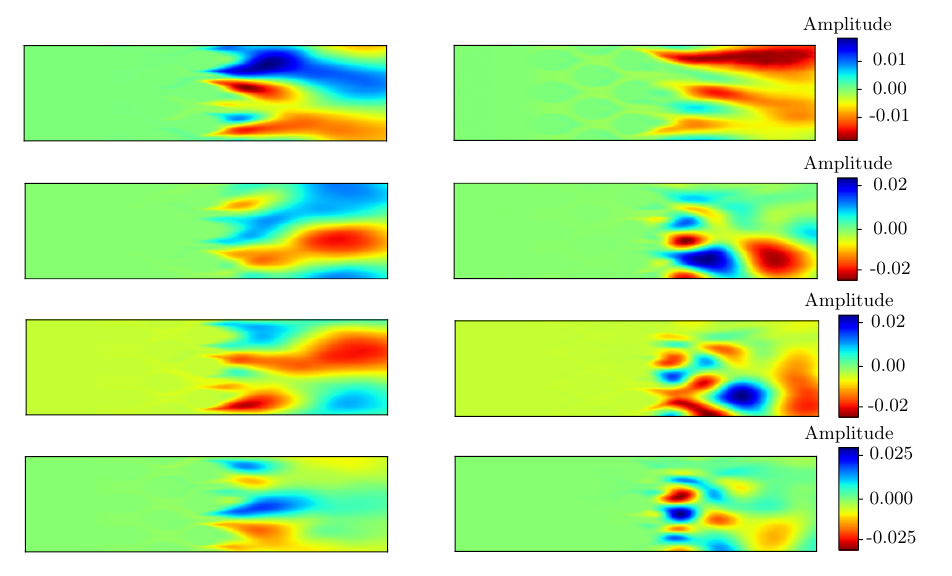}
    \caption{The four most dominant velocity modes, arranged from top to bottom. The left column corresponds to simulation results obtained from the lattice Boltzmann method, and the right column corresponds to results predicted using the 3D convolutional autoencoder for the simulation parametrized with $T^{\text{par, val}} = 750$ K.}
    \label{fig:modesVel}
\end{figure}

\begin{figure}[h]
    \centering
    \includegraphics[width=\textwidth]{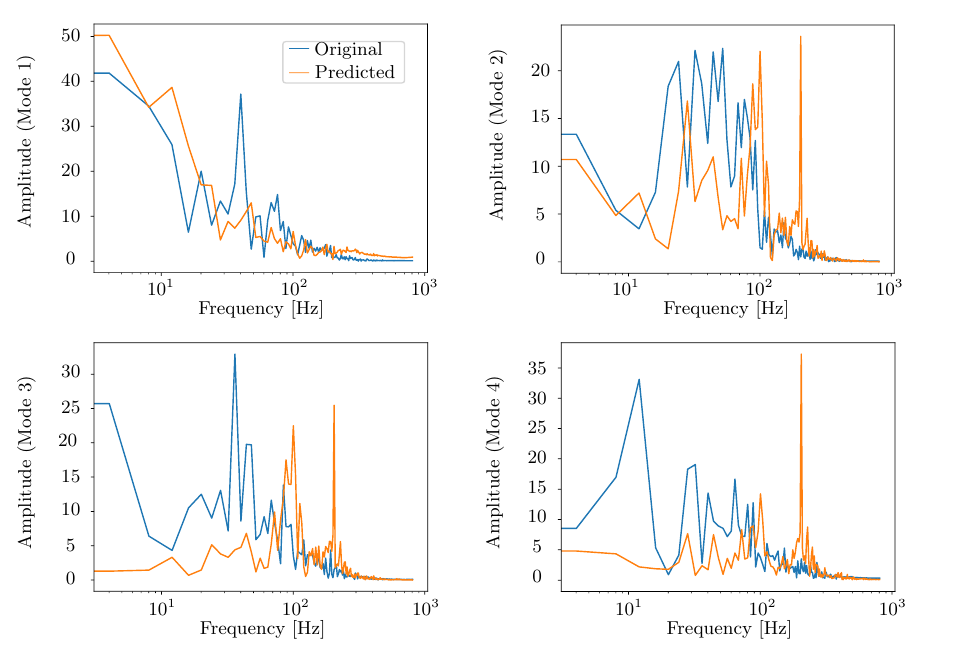}
    \caption{Power spectra for the temporal coefficients associated with the four most dominant POD velocity modes.}
    \label{fig:spectrumsvel}
\end{figure}

Similarly, we apply the POD method to the velocity field.
The four most dominant POD velocity modes are compared in Fig.~\ref{fig:modesVel}.
The results show that the near-upper-wall velocity stream, depicted in the mean of the predicted simulation in Fig.~\ref{fig:means}, is only partially visible in the first original POD velocity mode.
Upon a direct comparison of the modes, it becomes evident that both simulations exhibit the vortex shedding pattern in the downstream region, although with differing structures.
Specifically, the dominant pattern in the predicted solution consists of small-scale vortices, while the real solution features slightly larger ones.
For completeness, we illustrate in Fig.~\ref{fig:spectrumsvel} the spectra of the temporal coefficients associated with the shown POD velocity modes. However, a direct comparison is not feasible due to the distinct structure of the underlying POD subspaces.

In our case, the primary focus lies in predicting the transient and the post-transient solutions in the interstitial flow region between particles.
The downstream domain may not always require precise prediction in many practical applications. 
Instead, we can highlight the model's ability to accurately predict the flow field solution in the void space between particles in a bed configuration.    
Figures \ref{fig:temp_plots} to \ref{fig:spectrumsvel} show that our reduced model can predict the flow field solution for new simulation parameters well and that the dominant prediction errors are mainly located in the highly downstream region. 

\subsection{Comparison of the computational time}
While providing a precise comparison of the computational time required to evaluate the high-dimensional model against the reduced model is not possible due to the fine grid ($1800 \times 350$) used for the numerical solution, we directly compare the average computational time required to solve the high fidelity simulation \footnote{The high-fidelity model was solved on an Intel Xeon W-2255 CPU @ 3.7GHz 20} ($\approx 27365\text{ s}$) to that of the evaluation of the reduced model \footnote{The reduced model was trained and evaluated on an Intel Xeon Gold 6226R CPU @ 2.9 GHz} ($\approx 63\text{ s}$).
It is evident that a significant reduction in the computational time is obtained.

\section{Conclusion}
\label{sec:conclusion}
In this work, a reduced model for predicting flow fields in a bed configuration of hot spheres has been developed.
The model is data-driven and does not require access to the governing equations of the high-fidelity model.
Our work builds upon a previous study by \cite{Pant2021}, who use a 3D temporal convolutional autoencoder to learn a temporal sequence of 2D flow field snapshots to predict the temporal evaluation.
In our study, we have extended this framework by making it parametric, allowing the model to predict the solution for an unseen value of the parameter.
Therefore, our reduced model learns reduced order embeddings of high-dimensional dynamic systems and substantially reduces the computational time required to approximate the solution for a new value of the parameter.
Moreover, the framework presented here suggests using a post-processing non-trainable output layer to incorporate the available physical knowledge of the system into the prediction.
This is helpful to improve the overall quality of the results because the output of the model is stacked with the inputs at the next time step. 

In order to evaluate our model, we compared the original results of a simulation parameterized with a novel value of the parameter, which was not used during the training of the model, with the predicted solution using the parametric 3D temporal convolution autoencoder reduced model.
We emphasize that the prediction is made recursively and does not require supervision from the ground truth.
Significant CPU time and memory reduction have been obtained since the reduced model does not demand high dimensional discretization of the numerical domain to produce an accurate solution.
Results also showed that the model was able to accurately predict the flow field in the void region between particles, which is useful for many practical applications, including heat transfer analysis.
However, the performance of the reduced model in the downstream region is limited.
Despite this limitation, the overall vortex-shedding pattern in this region is captured. 

Even though the application of the presented framework to a densely packed bed configuration remains to be explored, this study shows the potential of using deep learning to build reduced models for the prediction of flow fields in the void space in a packed bed configuration.
In addition, the framework used in this study has potential applications in various engineering domains, such as heat exchangers.
Future work will focus on improving the prediction accuracy in the downstream region, considering highly packed beds and extending the framework to include more simulation parameters.

\section*{Acknowledgments}
Funded by the Deutsche Forschungsgemeinschaft (DFG, German Research Foundation) – Project-ID 422037413 – TRR 287. 

\section*{Appendix A. Proper Orthogonal Decomposition}
\label{appendix:POD}
Proper orthogonal decomposition (POD) is a technique introduced by \cite{Lumley1967} as a tool for extracting the dominant patterns in complex flow fields.
The core of POD lies in capturing the essential dynamics of a system through the decomposition of high-dimensional data into a reduced set of orthogonal modes called POD modes.

Consider a set of snapshots denoted as $\mathbf{q}(t_m) \in \mathbb{R}^N$, where $m=1,\cdots,M$.
The snapshot matrix $\mathbf{S} \in \mathbb{R}^{N \times M}$ is constructed by concatenating all the available snapshots across various time steps:
\[
\mathbf{S} = \begin{bmatrix}
\mathbf{q}(t_1) & \mathbf{q}(t_2) & \cdots & \mathbf{q}(t_M)
\end{bmatrix}.
\]
In the classical formulation introduced by \cite{Lumley1967}, the POD modes $\{ \boldsymbol{\phi}_1, \cdots, \boldsymbol{\phi}_r\} \subset \mathbb{R}^N$, where $r=\mathrm{rank(\mathbf{S})}$, are defined as eigenvectors of the correlation matrix $\mathbf{S} \mathbf{S}^T \in \mathbb{R}^{N \times N}$, i.e.,
\begin{equation}
\label{eq:eig}
\mathbf{S} \mathbf{S}^T \boldsymbol{\Phi} = \boldsymbol{\Phi} \boldsymbol{\Lambda}.
\end{equation}
The given eigenvalue problem is closely related to the singular value decomposition (SVD) of the snapshot matrix $\mathbf{S}$, which reads
\begin{equation}
\label{eq:svd}
\mathbf{S} = \mathbf{U} \boldsymbol{\Sigma} \mathbf{V}^T,
\end{equation}
where $\mathbf{U} \in \mathbb{R}^{N \times N}$ and $\mathbf{V} \in \mathbb{R}^{M \times M}$ are orthonormal matrices and $\boldsymbol{\Sigma} \in \mathbb{R}^{N \times M}$ is rectangular diagonal matrix with singular values $\sigma_1 \ge \sigma_2 \ge \cdots \ge \sigma_r$.
Using \eqref{eq:svd}, the correlation matrix can be written as:
\begin{align}
\label{eq:corrsvd}
\mathbf{S} \mathbf{S}^T = \mathbf{U} \boldsymbol{\Sigma} \mathbf{V}^T \mathbf{V} \boldsymbol{\Sigma} \mathbf{U}^T = \mathbf{U} \boldsymbol{\Sigma}^2 \mathbf{U}^T.
\end{align}
Right multiplying \eqref{eq:corrsvd} with $\mathbf{U}$ shows that the SVD of $\mathbf{S}$ solves the eigenvalue problem of the correlation matrix, i.e., extracts the POD modes
\begin{equation}
\label{eq:svdeqpod}
\mathbf{S} \mathbf{S}^T \mathbf{U} = \mathbf{U} \boldsymbol{\Sigma}^2.
\end{equation}

However, in many engineering applications, the direct computation of the POD modes becomes computationally intractable, especially when $\mathbf{q}_i$ results from a high dimensional discretization of the numerical domain, i.e., $N \gg 1$.
Therefore, \cite{Sirovich1987a} introduced the method of snapshots as a computationally efficient and tractable algorithm to extract the POD modes.

Once the POD modes have been determined, the quantity of interest $\mathbf{q}(t_m)$ can be approximated as a linear combination of the POD modes
\begin{equation}
\label{eq:lincomb}
\hat{\mathbf{q}}(t_m) \approx \sum_{i=1}^{K} a_i(t_m)\boldsymbol{\phi}_i,
\end{equation}
where $a_i$ is the temporal coefficient associated with the POD mode $\boldsymbol{\phi}_i$ at time step $t_m$.
If we choose $K = r$, the approximation is exact, i.e., $\hat{\mathbf{q}}(t_m) = \mathbf{q}(t_m)$, while choosing $K < r$ results in $\hat{\mathbf{q}}(t_m)$ being a low rank approximation of $\mathbf{q}(t_m)$ in the subspace spanned by $\{ \boldsymbol{\phi}_1, \cdots, \boldsymbol{\phi}_K\}$.
The accuracy of the approximation can be controlled such that
\begin{equation}
\epsilon = 1 - \frac{\sum_{i=1}^{K} \lambda_i}{\sum_{i=1}^{r} \lambda_i},
\end{equation}
is sufficiently small.
Here, $\lambda_i$ is the $i$-th eigenvalue of the correlation matrix $\mathbf{S} \mathbf{S}^T$ as indicated in \eqref{eq:eig}, which is equivalent to the square of the corresponding singular value of $\mathbf{S}$ as shown in \eqref{eq:svdeqpod}.

\bibliographystyle{plainnat}
\bibliography{references}

\end{document}